# Exploring the needs of Malay manuscript studies community for an e-learning platform

**Z. Zahidah**, **A. Noorhidawati** and **A.N. Zainab**

Digital Library research Group
Faculty of Computer Science & Information Technology,
University of Malaya, Kuala Lumpur, MALAYSIA
e-mail: zahidahz@perdana.um.edu.my; noorhidawati@um.edu.my, zainab@um.edu.my

**ABSTRACT**

*Philology studies are often associated with traditional methods of teaching and learning. This study explores the possibility of e-learning adoption amongst Malay manuscripts learning community. The Soft System Methodology (SSM) is used to guide the investigation. SSM emphasises on understanding the problem situations faced by Malay manuscript learning community and expresses the situations in rich pictures. The manuscript learning community comprises lecturers, students and researchers in the field of philology. Data were gathered from interviews, focus group discussions and observations. Academy of Malay Studies, University of Malaya is the case study setting, focusing on lecturers who teach and students who enrol in a philology course as well as doctoral students researching on manuscript studies. The findings highlight problems faced by the various stakeholders and propose solutions in the form of a conceptual model for a collaborative electronic platform to improve teaching and learning as well as utilizing digitized manuscript surrogates held in a digital library of Malay manuscripts.*

## INTRODUCTION

The main objective of digitization initiatives of cultural heritage such as ancient manuscripts in digital repositories or digital libraries is to preserve the fragile but valuable items and to make them widely accessible to the public. This has been evidenced by several digital library projects such as *MyManuskrip, the Digital Library of Malay Manuscripts* (Mohd Hilmi 2009; Mohd Hilmi and Zainab 2007), the *British Library Digital Catalogue of Illuminated Manuscripts, Better Access to Manuscripts and Browsing of Images (BAMBI)* (Rumpler and Calabretto 1999), *Euro-Mediterranean Union framework of Medievel Medicine (EUMME)* (Bozzi, Corradini and Tellez 2005), *an e-learning System for Greek Palaeography* (Drigas et al. 2005), and the *Bovary project* (Nicholas, Paquet and Heutte 2003, 2004).

Digital libraries however, have progressed from solely archiving digital materials to supporting other functions for their users such as being a platform for learning and research. In this context, digital libraries bring together collections, services and people to foster information and knowledge creation, dissemination, and sharing. Over the past few years digital libraries have been increasingly adopted into the learning and teaching environment since the progress of e-learning initiatives. There were reports about studies which utilized digital library in educational practices (Marshall et al. 2006; Abdullah and Zainab 2007; Yaron et al. 2008). In addition, Abdullah and Zainab (2008) in their research work on an integrated framework for strategic and architectural planning of digital library services, reported that digital library community not only desired a digital library where they could find resources, but also were also willing to be design partners and being part of the community within which they could contribute contents and communicate with others. This happen especially when digital libraries offer hard to get or rare materials to users who can use the resources to support teaching and learning in specialized studies such historiography, studies on cultural heritage, and philology. Access to rare collection through digital libraries makes it unnecessary for students studying philology or manuscript studies or academics teaching such courses to go physically to the library, sit in closely monitored rare collection rooms for hours to study the often fragile manuscripts. Digital libraries now make it possible for the manuscript studies community to work with the digitized surrogates.

Digital Library of Malay Manuscript known as *MyManuskrip* (available at http://mymanuskrip.fsktm.um.edu.my) is a digital repository of handwritten *Jawi* manuscript (adapted from Arabic alphabet) in Malay language. Malay manuscripts date as early as the 15th century and represents the historical records of some 500 years of Malay historiography, laws of the Malay Sultanate, Malay governance, ancient foreign policy, folk sciences, medicine, religion, beliefs, Malay literature and culture, Malay rites and rituals, astrological and folk literature such as *syair, gurindam, hikayat* (Malay form of sonnets). *MyManuskrip* is a digitization initiative developed mainly to preserve, organise, and make manuscripts accessible through the web for researchers, scholars and the public interested in reading, browsing or studying Malay Manuscripts.

*MyManuskrip* provides users with options to browse and search a single or cross collections of the digitized manuscripts and offer free access to university departments who offer courses on manuscript studies the use of the collections for teaching and learning, especially in transliterating works (Zainab, Abrizah and Hilmi 2009). However, *MyManuskrip* currently does not support an e-learning system to facilitate teaching, learning and doing research for the manuscripts community, especially in much needed areas such as for transliteration and annotation activities. Zainab, Abrizah and Hilmi (2009) found that one of the problems faced by the manuscript teaching and learning community is being unable to establish a dynamic collaborative platform amongst researchers and scholars or groups of manuscript studies students involved in joint transliteration work. Hence, a digital library that supports annotating and transliterating tools would help stimulate such social and educational networking. This paper investigates the processes and problems facing the manuscripts community in teaching, learning and researching Malay manuscripts. In order to understand the problems and suggest possible solutions, this study adopted the Soft System Methodology (SSM) that attempts to solve both soft and hard aspects of real-world problem situations (Checkland 1998; Achouri and Atkinson 2008).

## MANUSCRIPT DIGITAL LIBRARIES: A REVIEW

The development of digital libraries of ancient manuscripts is still considered scarce mainly because of the lack of expertise and limited budget to embark on manuscripts digitization process. In 2007, the Digital Library Research Group at the Faculty of Science Computer and Information Technology, University of Malaya, developed a digital library system, *MyManuskrip* or *Digital Library of Malay Manuscripts*. This initiative was started to provide a collaborative digital library environment for Malay manuscripts libraries to upload their digitized manuscript collections into a union repository and share resources (Zainab, Abrizah and Hilmi 2009). *MyManuskrip* (http://mymanuskrip.fsktm.um.edu.my/ Greenstone/images/html/mymanuskrip.htm) supports browsing, searching and online reading of the manuscripts deposited in the digital library. This digital library currently holds 179 titles from mainly two manuscript collections; the University of Malaya, Library and the Malay Documentation Centre, Dewan Bahasa dan Pustaka, Malaysia. The repository however could be enhanced by incorporating other features that could support teaching, learning and research using the available digitized manuscripts.

In the UK, The British Library's *Digital Catalogue of Illuminated manuscripts* (accessible at http://www.bl.uk/catalogues/illuminatedmanuscripts/welcome.htm) contains a number of different manuscript collections including a rich collection of medieval and renaissance manuscripts. The digital library provides access to images and information about its manuscripts to students, scholars, and the general public for reference. It provides a simple search function that enables searching by keywords and dates and advanced search using different combination of search options. Other services offered are static such as a thematic tour of various aspects of the British Library's illuminated manuscript holdings, with information about each manuscript, browsing and searching an illustrated glossary of terms and access to information about the scope and history of each collection.

The digital libraries as mentioned above mainly provide digital materials for users from their archives. It is necessary however, to incorporate other services in the digital libraries to support the teaching and learning process of manuscript studies by allowing individuals or group of collaborators to use the digital library for e-learning services. Such service, as reported by Calabretto and Bozzi (1998), is named *BAMBI* or *Better Access to Manuscripts and Browsing of Images*. BAMBI is a European digital archive of medieval manuscript collections which developed to provide a virtual platform for historian and philologist to browse and navigate the manuscripts in the collection (Rumpler and Calabretto 1999). in addition, BAMBI also provides a tool for philologists to write annotations, navigate between words of the transcription and match pieces of images in the numerated picture of the manuscript (Bozzi and Calabretto 1997; Calabretto and Bozzi 1998). This system used the standardized hypermedia language HyTime (Hypermedia/Time based Structured Language). Besides that, the *Euro-Mediterranean Union framework of Medievel Medicine* (*EUMME)* project is quite similar to BAMBI which focused on developing tools to facilitate users to read the manuscript collections it holds. EUMME aims to highlight and facilitate international collaboration among medico-pharmaceutical culture enthusiast to overcome the geographical borders, ideological and religious barriers among the community. EUMME is perceived as an effort to make the manuscripts available in different historical and philological environments which allows users to access homogenous digital archives in a collaborative environment (Bozzi, Corradini and Tellez 2005). Similarly, an *E-learning System for Greek Paleography* is an initiative to preserve antique documents and valuable manuscripts that focuses on Greek Paleography of the Byzantine era's collection of manuscripts that provides an interactive e-learning platform in order to value the collections in the educational practices. The electronic platform is part of the D-Scribe Project that provides features for recording process, manipulation, recognition and management of valuable Greek manuscripts and rare documents. This effort begins with the digitization of the manuscripts and later extends to provide an e-learning platform to support the process of teaching and learning Greek manuscripts (Drigas et al. 2005). The *Bovary* project on the other hand contains a collection of digitized manuscripts of Gustav Flaubert's manuscripts on Madam Bovary. This project initially started with collection of manuscripts without their transcriptions. It then progressed to include a text editor in order to help users to undertake transcription tasks more effective and help to produce a structured textual representation adapted to users' requirements. The text editor essentially provides an editing environment that integrates document analysis with interactive tools as well as making the original manuscripts collection available amongst researchers and specialists of Flaubert's works to foster collaboration (Nicolas, Paquet and Heutte 2003; 2004). Despite the reports of these projects, the digital libraries however cannot be accessed online.

## RESEARCH OJECTIVES

The objectives of this study are to:
  a) Examine the current processes of teaching, learning and researching Malay manuscripts;
  b) Identify issues, problems and difficulties face by Malay manuscript community while teaching, learning and researching Malay manuscripts;
  c) Propose a conceptual supporting platform that can improve the process of teaching, learning and researching Malay manuscripts.

This study aims to improve the current practice in teaching, learning and researching Malay manuscripts utilizing resources provided by *MyManuskrip* (*Digital Library of Malay Manuscripts*) using the Soft System approach.

## SOFT SYSTEM METHODOLOGY

Soft System Methodology (SSM) identifies problems in human activity using predefined seven stages (Figure 1). Human activity in this context refers to a collection of activities and relationship involving both predictable computer systems and unpredictable human behavior. SSM recognized that different individuals will have different perceptions and ways of working through problems and the proposed intervention should be acceptable to all parties. SSM is a generic model that has been used to investigate problem situations in a variety of organizations (Warwick 2008), digital library projects (Zainab, Abrizah and Hilmi 2009), information system development (Rose 2002; Somerville and Brar 2009), library management and information services (Delbridge and Fisher 2007; Delbridge 2008) and Information literacy (Nor Edzan 2007).

### Stage 1: Study Problem Situation

The research approach is a case study which allows the researchers to explore in depth an activity or a process undertaken by one or more individuals (Creswell 2009). The case(s) are bounded by place, time and activity, and the researchers collected detailed information using a variety of data collection procedures over a specified period of time. This method is anticipated to be suitable in situations where the case instances are small and focused as reflected by the very small number of institutions of higher learning offering manuscript studies as part of a philological course in Malaysia. In University of Malaya, there are 3 faculties that offer philological related courses namely Academy of Malay Studies, Academy of Islamic Studies and Faculty of Art and Social Sciences. Other universities that offer manuscript studies are Universiti Kebangsaan Malaysia, Universiti Putra Malaysia and Universiti Perguruan Sultan Idris. This study focuses on the Academy of Malay Studies, University of Malaya which is directly involved in teaching, learning and researching Malay manuscripts. The choice was based on convenience as the faculty members and the students involved are close at hand from the researchers' own faculty. This proximity is expected to facilitate the interviews and meetings with academic staffs and students.

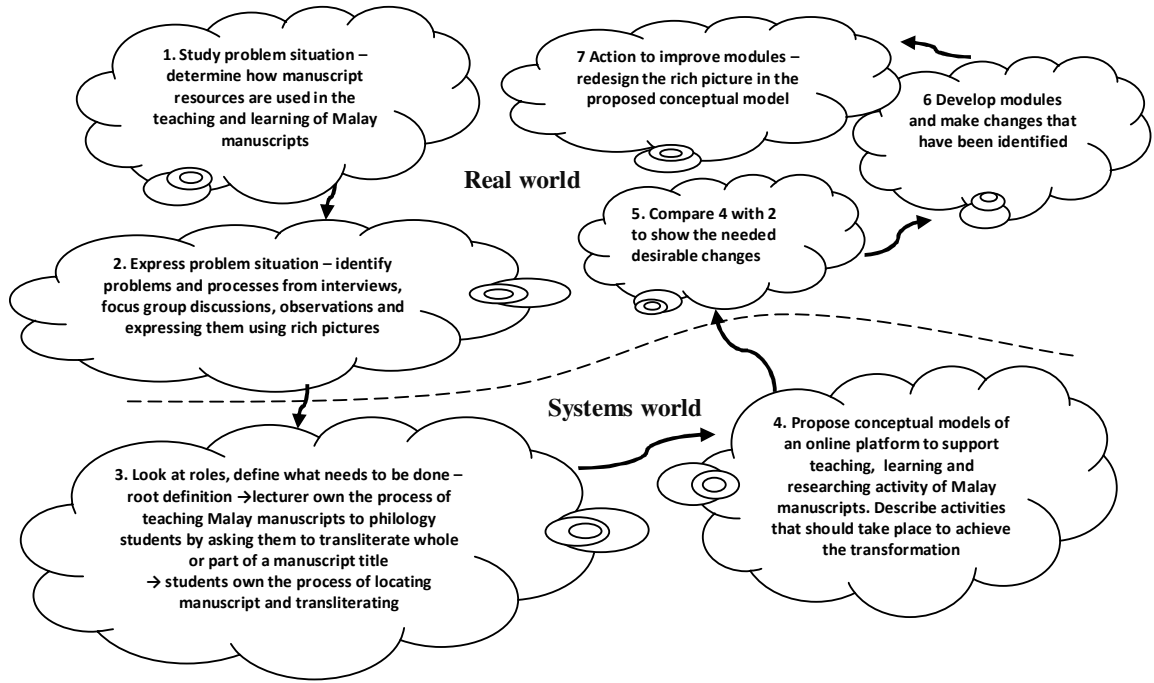

Figure 1: Application of SSM in Teaching, Learning and Researching Malay Manuscripts

In order to fully understand how the Malay manuscript users study Malay manuscripts, interviews, focus groups and observations were undertaken. According to Hancock (2002), qualitative interviews are semi structured or unstructured. An interview schedules which are tightly structured may results in the failure to investigate or explore the breadth and depth of the situation. Semi structured interviews work well when the interviewer has already identified a number of aspects he wants to explore. The interviewer can decide in advance the areas that need coverage but is open and receptive to unexpected feedback and information from the interviewees.

The interview sessions were conducted with lecturers who teach courses related to Malay manuscripts, students who enroll in Malay manuscripts courses and Malay manuscript researchers who were conducting research on Malay manuscripts. The main purpose is to investigate the current processes involved in teaching, learning and researching manuscript studies and to unveil issues related to it. For teaching and learning manuscripts, the interviews focused on three (3) lecturers who have had experiences handling manuscripts and also who are teaching Malay manuscript studies. In the Academy of Malay Studies, and the most related course that directly involves the use of Malay manuscript is "JEEA2303: Introduction to Philology". According to the Academy of Malay Studies Handbook (2008/2009), this is a core course for students enrolled at the Department of Malay Literature at the Academy. The handbook also explains that at the end of the course the students are expected to be able to explain the concept of philology, discuss works by local philologists in the Malay region and Europe, and be able to assess and examine philologists' works related to Malay manuscripts and apply the knowledge of philology to studying Malay manuscript that they have chosen as their study text for the course. In addition, students are expected to be able to read, understand and analyze the old Jawi scripts used in most Malay manuscripts. Students are also expected to transliterate (from old Jawi script to Romanized Malay) and translate (from old Romanized Malay to modern Malay) different manuscript versions and add their work as resources for future reference for users. Interview session is also conducted with students who are currently undertaking research on Malay manuscripts at the doctoral level.

Another data collection approach used was focus group. The focus group members are those who share common factors; in this study refers to those who are enrolled in a philology class. In this context the researchers observe the group's interactions to obtain "insights into the teaching and learning process that took place" (Hancock 2002). The focus group comprises 8 students of Academy of Malay Studies who have taken the Introduction to Philology course, who shared their experiences in handling and using Malay manuscripts, and the problems they faced during the study process. Participants were those who volunteered to be involved in the focus group. Members of the focus group were asked about their background especially on why they have chosen to follow the Malay manuscript studies and their suggestions on how to improve the study process. The first phase of the focus group discussions took up about one hour and forty-five minutes. All conversations were recorded, with the permission of the students being interviewed. After the session, the recordings were transcribed into computer files. Care was taken to assure the respondents of the confidentiality of their responses. Once the transcriptions were completed, copies were printed out to be verified by the respondents themselves. All of the focus group transcripts were read by the researchers and coded into themes to facilitate data analysis.

Finally, the observation approach was used to get views about the process of teaching and learning Malay manuscripts. During the second semester of 2009/2010 session the researchers observed the "JEEA2303: Introduction to philology" classes for 14 weeks (one semester), from 12 January 2010 until 6 April 2010. The classes were separated into 2 sessions which comprised lectures and tutorials. The lecturer who taught the classes suggested that the researchers observe only the tutorial classes as it was in these sessions that the manuscript study discussions and practices took place. In these tutorial sessions the researchers merely observed without participating in the class activities. This helped verify remarks made during the interviews and focus group discussion sessions. The total numbers of students in the tutorial classes observed was 46 students who were divided into two groups. During each tutorial class, the class discussions and activities were recorded and notes were taken to describe what took place in class. The thematic analysis was chosen as the data analysis method. This type of analysis is highly inductive, as the themes emerge from the data and are not imposed upon by the researchers.

### Stage 2: Express Problem Situations Using Rich Pictures

This stage depicts the processes and problems identified from the interview, focus group and observation sessions in a rich picture as shown in Figure 2. Findings from the three data collection approaches were extracted and collated according to three different types of stakeholders, namely (i) three (3) lecturers (L1, L2 and L3) who are known to have higher knowledge on Malay manuscripts, and were handling and teaching Malay manuscript studies during the time of the study, (ii) eight (8) students (S1, S2, S3, S4, S5, S6, S7 and S8) who admitted having very basic level of knowledge about Malay manuscripts, and were registered or had undergone an introductory course in Malay Philology and (ii) three (3) researchers (R1, R2 and R3) who are researching on Malay manuscripts and who were registered for doctoral programme during the time of the study.

The findings from this study encompassed merging data from the coding process, which involved taking text data, segmenting sentences (or paragraphs) into categories, and labeling those categories with terms, often a term in the actual language of the participant. This process is recommended by Creswell (2009) in order to break the responses obtained into themes. The descriptions aims to highlight the processes of teaching and learning Malay manuscripts presently practiced the supporting resources available and the use of technologies in the teaching and learning process.

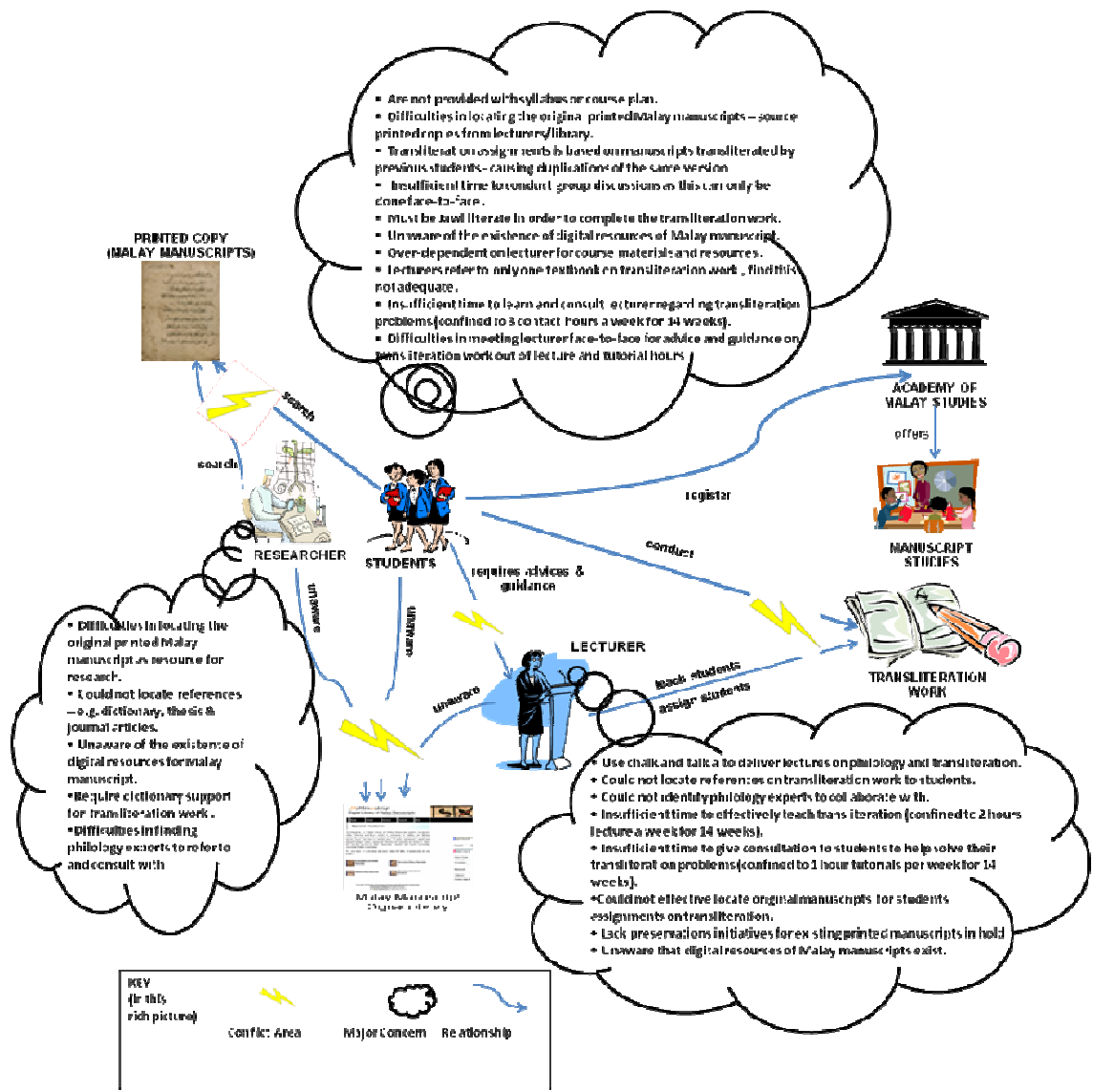

Figure 2: Problems Expressed by Students, Lecturers and Researchers in a Rich Picture

## (a) Lecturers Teaching Manuscript Studies

The responses from the lecturers were collated from three interviews and one observation session. The following sections indicated the responses listed under three broad themes, namely Teaching and Learning Manuscripts Currently Practiced (T1), Supporting Resources for Manuscript Studies (T2) and Technology Use and Acceptance (T3).

### (i) Teaching and Learning Manuscripts Currently Practised (T1)

During the interviews, the lecturers described the main process of teaching Malay manuscripts, which mainly involved transliteration. Transliteration is the process of mapping from one system of writing into another, word by word, or ideally letter by letter and in this class, students were required to transliterate the old Jawi scripts into roman Malay without changing phonetics and grammar. One of the lecturers (L1) explained, "*There are fixed method that had been taught in the class, all these methods can be found in the reference book. The methods are transliteration, transcription and annotation. But all these methods are done manually*". When asked about the process of learning a Malay manuscript, L2 said that, "*Firstly, students will do the transliteration and later the analysis. If they did the wrong transliteration, they will later produce a wrong analysis. Usually, I will ask them to see the librarian of the National Library to help them to improve their work*". Another remarked, "*As a philologist, all you have to do is to know how to read manuscript and do the analysis*" (L3).

### (ii) Supporting Resources for Manuscript Studies (T2)

Supporting resources in this context refers to time allocation for manuscript studies, availability of printed original manuscripts, problematic type of students following manuscript study course, the availability preservation initiatives and the availability of experts.

- ▪ *Time allocated to manuscript studies*

Lecturers indicated that manuscript studies is covered in only one course that is "Introduction to Philology" as L1 described "*In every semester, there are 14 weeks for lecturers to do the teaching, and this is problematic for us as we have to optimize this limited time*". This indicates that the lecturer felt the time allocated to manuscript studies is insufficient.

- ▪ *Resource Support*

A lecturer (L3) described the problems his students faced in getting manuscripts: "*The students are always complaining that it is hard to find Malay manuscript resources. This year, our manuscript club is gathering all the manuscripts we can find and keeping them at the Academy. However, students still complain that it is hard to get access to them and the number of manuscripts is also very small. To solve this, I don't have any choice, but to make them refer to the printed version of manuscripts which is a simplified version done by other researchers*". This infers that instead of working on original manuscripts students were asked to use the printed or edited version of Malay manuscripts, which means that they have to work on other people's work. This results in the duplication of transliteration works. Whether this method simplifies the learning of transliterations need further investigation. The problem of access is also indicated by this Academy of Malay Studies lecturer, "*To find Malay manuscripts that have never been used by previous students is somehow a problem to the current students. There is a lack of Malay manuscripts resources that are easily* *available*" (L1). The feedback indicated that both lecturers and students were finding problems in locating and accessing manuscripts for transliteration assignments.

■ **Students Signing-up for Philology Course**

Lecturers indicate that there are students who do not know how to read Jawi but were taking the "Introduction to philology" course (which is a core course). Lecturers felt sorry for these students and could do little to help since they felt that teaching Jawi is too basic to be taught at the University level. This infers that lecturers have problems when some students signing-up for their course are not able to read Jawi.

■ **Initiatives to Conserve Manuscripts**

To save and protect old manuscripts needs initiative and budget. One lecturer (L3) indicated, "*Like I said earlier, the people from Arts Department need to have the initiative to learn and preserve (Jawi manuscripts), but I never see that. I can see other people from other Departments like me who are more interested compared to them*".

Another (L2) remarked, "*Like in this university, currently there is no 'special' place to keep the original manuscripts. Manuscripts need to be located in a room with a suitable temperature and people cannot simply enter the room, usually they need to wear mask and gloves in order to protect the manuscripts that are getting older and fragile. There are regulations on this matter but most people are not aware of it*".

"*Currently, all the manuscripts are located in the main library, but there is no one there to take care of the manuscripts. For me, I can build a room for manuscripts, but I need finances which would not exceed 100 thousand ringgit*" (L3). These remarks indicate that lecturers believed that there was no initiative to conserve Malay manuscripts. They are aware for the need to protect manuscript collections but the amount mentioned that is required for the conservation work is actually unrealistically low.

■ **Experts in Manuscript Studies**

One lecturer at the Academy of Malay Studies (APM) opined, "*In APM, there is no philology expert*" (L3). Another lecturer (L2) indicated that "*At APM, there is a society that is responsible for promoting Malay manuscripts and literary works and they are currently progressing by making plans to protect old manuscripts. Some lecturers in APM are concerned about saving the old manuscripts. On their own initiative, they microfilmed and digitized the manuscripts*".

**(iii) Technology Use and Acceptance (T3)**

The lecturers interviewed believed that automation projects makes students lazy and learning in an electronic environment is not enough to be a philologist. One lecture (L3) expressed his concern about making manuscript copies available on the Internet and argued over the ownership of such manuscripts and also the legality of making manuscripts accessible on the web. He commented, "*Actually, this electronic thing will make people lazy. For philologist, they have to know not only the physical aspects of the manuscripts, but they need to feel and touch the original manuscript. Also, there is no guarantee that all the information on the Internet is true.*"

**(b) Students Following Manuscript Studies Course**

The data from students were collated from the interview sessions, focus groups and also the observations. Their responses are described below in the 3 broad themes as mentioned earlier.

**(i) Teaching and Learning Manuscripts Currently Practised. (T1)**

From observations made during tutorial classes, the student found that the lecturer would directly start the class by introducing the subject and explaining the meaning of philology. The explanation was delivered without using any supporting material such as slides presentation. The resource used by the lecturer is a reference book and sometimes he wrote down the important terms on the white board. One of the student in the focus group session remarked about this, "*Currently, in philology class, there is no syllabus provided by the lecturer in the beginning of the class such as the objectives of the subject and the weekly plan*" (S1). When asked about the class supporting materials, one of the students said that, "*We only use the reference books provided by the lecturer and the lecturer uses only one reference book*" (S1). This seems to imply that students in the philology course were very "lecturer-dependent". Nobody complained directly to their lecturers indicating that although they may not be satisfied they believe that this is the normal method of learning manuscript studies.

**(ii) Supporting Resources for Manuscript Studies (T2)**

▪ *Time Allocated to Manuscript Studies*

In the tutorial classes every week, students will present the updates on the progress of their assignment. During these sessions each student share and discuss problems faced with their lecturer and peers in class. One student explained, "*The duration of 1 semester to learn Malay manuscript is not enough. Usually we only have time to have discussions with lecturers during the tutorial session which is only 2 hours per week and usually not all students get a chance to present and seek for lecturer's comments*" (S2). Concurring with their lecturers, students also expressed that the time allocated for the course is insufficient.

▪ *Resource Support*

Another issue students brought up is about the lack of Malay manuscript resources. "*There are no specific resources where we can easily locate Malay manuscripts. Besides the main library, we usually get the manuscript from our own collection at home, but we need to show our lecturer first and get his approval before we proceed with our research on that manuscript*" (S1).

▪ *Initiatives to Learn about Manuscripts*

The students believed that their generation should have more initiative to learn Malay manuscripts in order to understand the need to protect this heritage. One student said, "*We are lucky to be in the philology class and got a chance to learn Malay manuscripts. I think it is good to open this subject to all students from any faculties so that at least it helps us learn and know the creative works of our previous ancestors*" (S5).

▪ *Expertise*

In class, students refer problems or bring issues to their lecturer and when asked about other resources, S2 answered, "*Currently our main reference is our lecturer. We don't have connection with other lecturers or manuscript experts*". Another student added, "*During the class, the discussion is limited only between students and lecturer. So, everything is based on the lecturer's knowledge*".

Mainly, lecturer would suggest one reference book for students to refer to; students may get other information during the class by jotting down notes from what had been delivered by the lecturer. One student commented, "*The reference book that are being used in the* *class is not enough, it gives students only the basic knowledge about Malay manuscripts and philology. Students also need to refer to other resources such as theses" (S4).*

### (iii) Technology Use and Acceptance (T3)

All students indicated being frequent users of the computer and internet, such as the social networking web. One student (S6) commented, *"When I was involved in the exchange students programme with a student from China, we are connected through a blog and I find it is very helpful for us to work together at anytime and anywhere"*. These findings indicated that there is no problem for the students to accept technologies in their daily activities. Another student remarked*, "It's good to have such a repository [proposed by the researchers] on the web where we can share everything related with the subject, so that we do not rely only on the 4 hours time a week to do our research on Malay manuscript"*(S8). The students taking the Philology course like most students of their age seems ready to accept and use ICT to communicate and learn the subject.

## (c) Researchers

The data from researchers sampled were collated from the interview sessions. Their responses are also described below in the 3 broad themes as mentioned earlier.

### (i) Current Practice of Manuscript Research Activity (T1)

From the researchers' views, besides being able to read the Jawi script, it is important to have skills in transliteration as one researcher remarked, *"Transliteration is the most important process where the researcher needs to convert the old Jawi script to romanise Malay. It is very important for the researcher to know how to read Jawi script and have the skills in understanding the old Jawi script that is different when compared to the modern Jawi script that are being taught in schools today"* (R1).

### (ii) Supporting Resources for Manuscript Studies (T2)
- *Resource Support*

It is clear that some respondents were not aware of other available resources beside the print copies. One researcher commented, *"I have never heard about online resources in manuscripts collection before. All this while, I use the research material from books and thesis that are available in libraries"* (R2). This is surprising as there are manuscript digital libraries available on the Web.

- *Expert Support*

Researchers sampled in this study were mainly postgraduate students. Their sources of reference are usually lecturers and experts from various universities. One researcher said, *"I always keep myself in touch with expertise in my research area from other universities in order to keep myself updated and get information related to my research."* (R1)

Beside reference books, they also refer to theses as their main research resource, as further confirmed by R1, *"Yes, the collection of theses from other researchers is very important as my research resources. I always get it from the faculty library or directly from the owner of the thesis"*. This indicate the researchers are not information literate in terms of sourcing out manuscript information as versions of Malay manuscripts are available in various libraries in the world. This applies also to Malay manuscript experts who are not confined only to Malaysians.

**(iii) Technology Use and Acceptance (T3)**

Researchers from the Academy of Malay Studies indicated that they seldom use computers or the Internet while doing their research. They felt that technology usage is unnecessary. They would use computers only for word processing their chapters, as one researcher remarked, "*I seldom use PC in my daily research. I just use it for writing, most of the research sources are in hard copies and I believe that all the researchers in manuscripts studies area do research this same way*" (R3).

However, when asked about using technologies to support their research, they have no problems in accepting it as a supporting tool. As indicated by one researcher, "*Besides my own experience, I also use dictionaries to help me in transliteration. There are quite a number of dictionaries in the market, but so far I think "Kamus Melayu Klasik" is the best dictionary for Malay manuscript studies. If this dictionary can be digitized, it will be very helpful*" (R2).

**Stage 3: Look at Problems using CATWOE analysis**

Stage 3 looks at the problems as illustrated from the rich picture and propose solutions for improvement by delineating a "root definition" using the mnemonic CATWOE analysis. The root definition is essentially a hypothesis about the relevant situation and improvement to it that might help to resolve the problem situation. Therefore the working root definition for the Malay manuscript community is:

> "An instructor/lecturer owned an operated system to provide a conducive medium for teaching, learning and researching, by giving them assignment through potential e-learning tool such as transliteration and annotation tool in order to support collaboration work among researcher/lecturer/students and people who are involved in Malay manuscript studies through a digital environment".

Table 2 details the elements of the root definition through a CATWOE analysis.

Table 2: The Elements of the Root Definition and the CATWOE analysis

| Element of CATWOE | Description | CATWOE analysis for this research |
|---|---|---|
| Customers | Who are the victims or beneficiaries of the transformation? | Manuscripts lecturers<br>Manuscripts students<br>Manuscript researchers<br>The public accessing Malay manuscripts Collections |
| Actors | Who makes the transformation happen? | The manuscript librarians in charge of governing the Malay manuscripts collections |
| Transformation | What are the inputs and (transformed) outputs? | Current unconducive environment of teaching, learning and researching Malay manuscript to an accommodating environment |
| Weltanschauung | What makes the transformation meaningful in context? | A conducive medium for teaching, learning and researching Malay manuscript that support interactive and collaborative in a digital environment |
| Owners | Who could stop the transformation process? | FCSIT or Collaborating Malay manuscripts repositories in Malaysia |
| Environmental Constrains | Which elements outside the system are taken as given? | Technology Use and Acceptance<br>Collaboration between researchers |

## Stage 4: Conceptual Model

A conceptual model is developed to show how various activities of different stakeholders in the current system are related to each other based on the root definition as shown in the Figure 3. The conceptual model demonstrates how technology can help solve the problems expressed by the Malay manuscript communities by mapping it to a proposed system that can be used by the three different stakeholders as shown in Table 3.

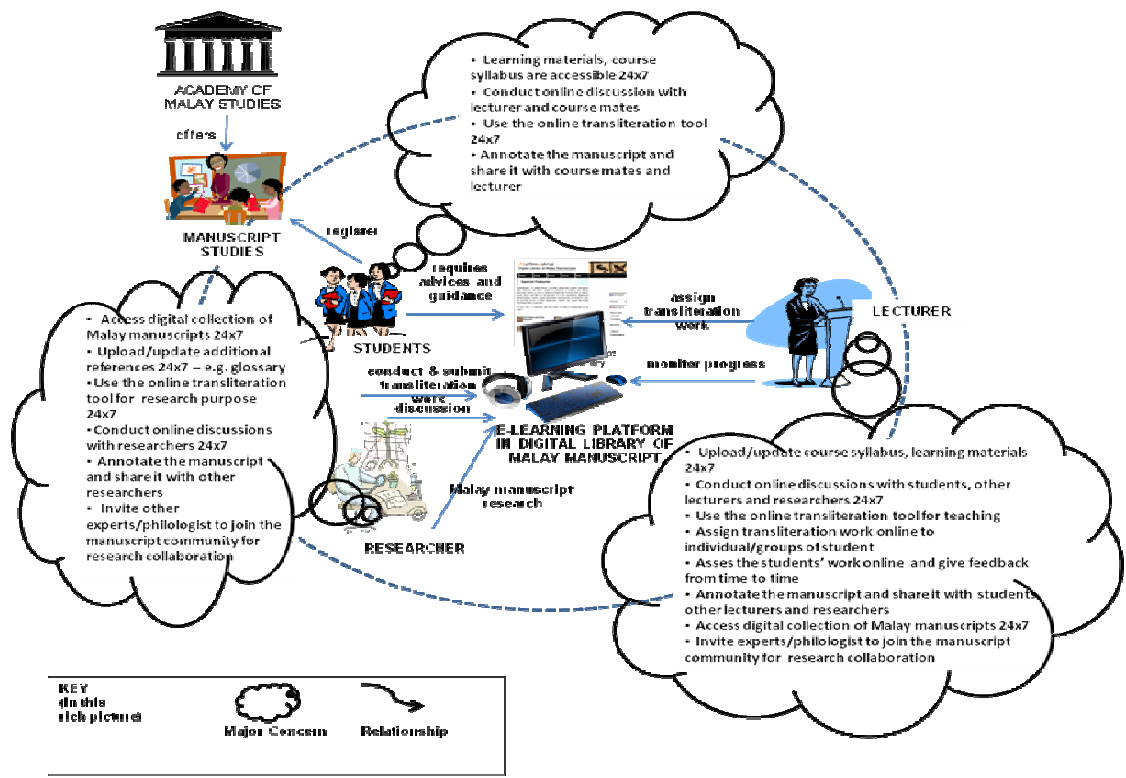

Figure 3: Conceptual Model Representing the Solution for Problems

## Stage 5: Comparing the Current Situation with the Proposed Solution

This stage compares the conceptual model with the current system that identifies the desirable and feasible changes needed as shown in Table 3.

## Stage 6 & 7: Feasible, Desirable Changes Needed and Action to Improve the Problems

This stage implements all the changes that had been identified. The rich picture was redesigned into the conceptual framework as shown in Figure 3 which proposed the process of teaching, learning and researching Malay manuscripts to be done in a digital environment in order to solve the issues and problems that had been identified in the current system. The details on the desirable changes that should be undertaken are being analyzed and will be followed by the actual implementation of the digital library. The provision of an online platform to support the teaching, learning and research activities together with the use of resources from *MyManuskrip*, is generally welcomed by philology lecturers, students and researchers at the Academy of Malay Studies, University of Malaya. "*We are glad to hear that there is such thing like MyManuskrip which is can help my* *students looking for the collection of Malay manuscripts. We hope that you are going back here with the e-learning system and going to be used in my class*" (L1). Specifically, the modules proposed should support transliteration and annotation activities, use current digital manuscript resources from *MyManuskrip*, allow lecturers to upload their syllabus and teaching resources, support interactivity in terms of student-lecturer, student-student discussions; the uploading of transliterated assignments as a collection in MyManuskrip; and the provision of a dictionary of terms to assist manuscript researchers. Generally, the proposed system should be able to support discussions and the sharing of knowledge while promoting the use of Malay Manuscript widely.

Table 3: Comparison between the Current Situations with the Proposed Solution

| Stakeholders | Current Situation | Proposed Solution |
|---|---|---|
| Lecturer | <ul><li>Use chalk and talk as delivery lecture on philology and transliteration.</li><li>Could not locate references on transliteration work.</li><li>Could not identify philology experts to collaborate with.</li><li>Insufficient time to teach transliteration (confined to 2 hours lecture a week for 14 weeks).</li><li>Insufficient time to give consultation to students to help solve their transliteration problems (confined to 1 hour tutorials a week for 14 weeks).</li><li>Could not locate original Malay manuscript for students assignments on transliteration.</li><li>Lack preservations initiatives for existing printed manuscripts.</li><li>Unaware of the existence of digital resources on Malay manuscript.</li></ul> | <ul><li>Experience teaching in an electronic environment.</li><li>Upload/update course syllabus, learning materials 24x7</li><li>Conduct online discussions with students, other lecturers, experts (in Malaysia and abroad), and researchers 24x7.</li><li>Use the online transliteration tool for teaching.</li><li>Assign transliteration work online to individual/groups of student.</li><li>Asses students' work online and give feedback from time to time.</li><li>Annotate manuscripts and share with students, other lecturers and researchers.</li><li>Access digital collection of Malay manuscripts 24x7.</li><li>Invite experts/philologist to join the manuscript community for research collaboration</li></ul> |
| Students | <ul><li>Are not provided with a proper syllabus or course contents.</li><li>Difficulties in locating the original printed Malay manuscript – source printed copy from lecturer/library</li><li>Transliteration assignments is based on manuscript transliterated by previous students – causing duplication of the same title.</li><li>Insufficient time to conduct group discussion as this can only be done face-to-face</li><li>Must be Jawi literate in order to complete the transliteration work</li><li>Unaware of the existence of digital resources of Malay manuscript</li><li>Students are over-dependent to lecturer for course materials and resources</li><li>Lecturers refer to only one textbook on transliteration work, find this inadequate.</li><li>Insufficient time to learn and consult lecturer regarding transliteration work (confined to 3 hours lecture & tutorial sessions a week for 14 weeks)</li><li>Difficulties in meeting lecturers face-to-face for advice and consultation on transliteration out of lecturer's tutorial hours</li></ul> | <ul><li>Learning materials, course syllabus are accessible 24x7 .</li><li>Conduct online discussion and consultation with lecturer and course mates.</li><li>Use the online transliteration tool 24x7 .</li><li>Annotate the manuscript and share it with course mates and lecturer s.</li><li>Browse collection of digitized manuscripts in digital libraries in Malaysia and abroad.</li><li>Use new digital surrogates of manuscripts from digital libraries and repositories for transliteration assignments.</li></ul> |
| Researcher | <ul><li>Difficulties in locating the original printed Malay manuscripts for research</li><li>Could not locate references to refer to – e.g. dictionary, thesis & journal papers</li><li>Unaware of the existence of digital resources for Malay manuscript</li><li>Require dictionary support for transliteration work</li><li>Difficulties to find philology experts to refer to and consult.</li></ul> | <ul><li>Access digital collection of Malay manuscripts 24x7</li><li>Upload/update additional references 24x7 – e.g. glossary</li><li>Use the online transliteration tool for research purpose 24x7</li><li>Conduct online discussions with researchers 24x7</li><li>Annotate the manuscript and share it with other researchers</li><li>Invite other experts/philologist to join the manuscript community for research collaboration</li></ul> |

## CONCLUSION

In summary, this study has shown that by employing SSM, the process and issues related to teaching, learning and researching Malay manuscripts were captured and depicted in a rich picture based on three different views of stakeholders:

(i) lecturers mainly used the "talk and chalk" approach when delivering their course; they referred students to one reference source throughout the course (Ghani and Zakaria 2009); they indicated that students must be able to read Jawi scripts to follow the course; the main assignment involved transliterating work from old Jawi scripted text into modern romanise Malay; and they would like to have access to more Malay manuscript texts that can be utilised by their students for their assignments.

(ii) students confirmed that their main assignments involved transliterating work; their main source of reference was their lecturer; they were unaware of the existence of a digital library of Malay manuscripts; and were acceptable to accessing online manuscript resources and being able to hold online discussions with their lecturers and peers about their assignments.

(iii) researchers used printed text for their research; they used the computers mainly to word-process; and felt that online dictionary on classical Malay would help them in their research.

In order to solve the problem situation as illustrated in the rich pictures a proposition of the relevant solutions were defined through a root definition which was mainly to provide a conducive medium for teaching, learning and researching Malay manuscript that is highly interactive and supported collaboration work through a digital environment. Hence features of the proposed solution were defined and conceptual model was developed to show interaction between different views of stakeholders and described the system processes to achieve the desired transformation. For example, allowing students to conduct a group discussion with no time limit and place, is the proposed improvement compared to the current system where students were confined to the I hour tutorial sessions for 14 weeks to discuss their progress and problems faced with peers and lecturers during the transliteration assignment. Finally, the rich picture is redesigned into a new conceptual framework which proposed the process of teaching, learning and researching Malay manuscripts to be done in a digital environment in order to solve the issues and problems that had been identified in the current system. The details on the desirable changes that should be undertaken is being analyzed and will be followed by the actual implementation of the e-learning system linked to a digital library (*MyManuskrip*), to support the teaching, learning and research activities utilising resources from *MyManuskrip*. This will ease locating and access problems as highlighted by Ding (1987) and Zuraidah (2008) and all stakeholders in this study. The digital environment would also support transliteration and annotation activities, allow lecturers to upload their syllabus and teaching resources, support interactivity in terms of student-lecturer, student-student discussions; the upload transliterated assignments as a collection in *MyManuskrip*; and the support of glossary of terms to assist manuscript researchers. In general, the proposed system should be able to support collaborations and the sharing of knowledge while promoting the use and study of Malay manuscripts widely.